\documentclass[a4paper,12pt]{article}
\pdfoutput=1
\usepackage{amsmath}
\usepackage{amsfonts}
\usepackage{amssymb}
\usepackage{graphicx, rotating}
\usepackage{epsfig}
\usepackage{latexsym}
\usepackage{graphicx}
\usepackage{xcolor}
\usepackage{amsmath,bm,amssymb}

\usepackage[normalem]{ulem}

\usepackage{color}
\usepackage[english]{babel}
\usepackage[colorlinks,allcolors=blue]{hyperref}
\usepackage[numbers,sort&compress]{natbib}

\newcommand{\Slash}[1]{{\ooalign{\hfil#1\hfil\crcr\raise.167ex\hbox{/}}}}

\newcommand{\beq}{\begin{equation}}  \newcommand{\eeq}{\end{equation}}
\newcommand{\bef}{\begin{figure}}  \newcommand{\eef}{\end{figure}}
\newcommand{\bec}{\begin{center}}  \newcommand{\eec}{\end{center}}

\def\({\left(}
\def\){\right)}

\begin{document}
\begin{titlepage}
\begin{center}
\setcounter{footnote}{0}
\setcounter{figure}{0}
\setcounter{table}{0}

\hfill    \\

\vspace{1.0cm}

{\Large\bf 
	Probing the high temperature symmetry breaking with gravitational waves from domain walls}

\vskip .75in

{ \large Xiu-Fei Li}

\vskip 0.25in

\begin{tabular}{cc}
& {\em School of Physical Science and Technology, Inner Mongolia University, }\\
& {\em Hohhot 010021, China}\\
& {\em xiufeili@imu.edu.cn}\\[.3em]

\vspace{12pt}
\vspace{0.1cm}

\end{tabular}

\vspace{0.5cm}
\abstract{
The symmetry can be broken at high temperature and then restored at low temperature, which is the so-called \emph{high temperature symmetry breaking}. It often appears in some  theories such as the high scale electroweak baryogenesis mechanism. In this paper, we probe the high temperature $\mathbb{Z}_2$ symmetry breaking with  gravitational waves (GWs) from domain wall annihilation. We first introduce a scalar  with $\mathbb{Z}_2$ symmetry and few of singlet fermions that interact with  scalar through a five-dimension operator. This can lead to the scalar potential has a non-zero minimum at high temperature. At the early stage, the scalar is pinned at symmetric phase due to the large Hubble fraction.  When the scalar thermal  mass becomes comparable to the Hubble parameter, it can quickly roll down to the minimum of potential. Then the $\mathbb{Z}_2$ symmetry is spontaneously broken and the domain walls will form. With the decrease of temperature, $\mathbb{Z}_2$ symmetry will be restored. We find that if domain walls are formed at $\mathcal{O}(10^{9})~ \rm GeV$, the GW produced by domain wall annihilation is expected to be observed by BBO, CE and ET. In addition, we also discuss the relationships between this scenario and NANOGrav signal. 
}

\end{center}
\clearpage

\noindent


\end{titlepage}
\setcounter{footnote}{0}
\setcounter{page}{1}

\section{Introduction}              
The first direct detection of  gravitational wave (GW) by LIGO \cite{LIGOScientific:2016aoc} in 2015 marked the beginning of GW astronomy. Most of the GWs come from violent astronomical processes such as black hole and neutron star binaries. But there are also many potential cosmological sources for the stochastic GW backgrounds, such as inflation \cite{Bartolo:2016ami}, reheating followed by inflation \cite{Caprini:2018mtu}, cosmic string \cite{Auclair:2019wcv} and the first order cosmological phase transitions \cite{Caprini:2015zlo,Caprini:2019egz}. The GW propagates freely once they are generated, which provides a powerful tool for exploring the early Universe.

Very recently, the North American Nanohertz Observatory for Gravitational Waves collaboration (NANOGrav) has released the data collected in the first $15{\rm\,years}$ of activity~\cite{NANOGrav:2023gor, NANOGrav:2023icp, NANOGrav:2023hfp, NANOGrav:2023ctt, NANOGrav:2023hvm}, which shows the compelling evidences of a stochastic GW background around $f\sim 10^{-8} ~\rm Hz$. Similar resluts have also been independently reported by the Chinese Pulsar Timing Array (CPTA)~\cite{Xu:2023wog}, the European Pulsar Timing Array   (EPTA)~\cite{Antoniadis:2023lym, Antoniadis:2023puu, Antoniadis:2023ott, Antoniadis:2023aac, Antoniadis:2023xlr, Smarra:2023ljf}, and the Parkes Pulsar Timing Array (PPTA)~\cite{Zic:2023gta, Reardon:2023zen, Reardon:2023gzh}. There are many papers have studied the possible origin or implications of this observation from the point of view of  inflation \cite{Vagnozzi:2023lwo}, first order phase transitions \cite{Madge:2023cak,Megias:2023kiy, Fujikura:2023lkn, Athron:2023mer, Addazi:2023jvg, Han:2023olf, Zu:2023olm, Yang:2023qlf, Li:2023bxy, Xiao:2023dbb,Ghosh:2023aum,Gouttenoire:2023bqy}, cosmic strings \cite{Ellis:2023tsl, Wang:2023len, Bian:2023dnv,Lazarides:2023ksx,Antusch:2023zjk,Buchmuller:2023aus}, domain walls \cite{Kitajima:2023cek, Bai:2023cqj, Gouttenoire:2023ftk, Blasi:2023sej,  Lu:2023mcz,Barman:2023fad,Du:2023qvj,Babichev:2023pbf}, axions or axion-like particles \cite{Yang:2023aak, Guo:2023hyp}, supersymmetry \cite{Murai:2023gkv,Du:2023qvj}, fuzzy dark matter~\cite{Anchordoqui:2023tln}, supermassive black hole
binaries (SMBHB) \cite{Ellis:2023dgf,Shen:2023pan},  primordial black holes \cite{Franciolini:2023pbf,Liu:2023ymk} and so on. Here we focus on the domain wall annihilation followed by the high temperature $\mathbb{Z}_2$ symmetry breaking.\footnote{Generically, the means of ``~high temeprture symmetry breaking~" is that the temperature of symmetry breaking is higher than that of symmetry restoration, even if the energy scale of symmetry breaking is not very high. } 

The high temperature symmetry breaking in the early Universe is an interesting concept \cite{Weinberg:1974hy,Mohapatra:1979qt,Fujimoto:1984hr,Dvali:1995cj,Salomonson:1984rh,Bimonte:1995sc,Dvali:1996zr,Orloff:1996yn,Gavela:1998ux,Ahriche:2010kh,Espinosa:2004pn,Bajc:1998jr,Agrawal:2021alq}, which has aroused great concern \cite{Meade:2018saz,Baldes:2018nel,Glioti:2018roy,Matsedonskyi:2020mlz,Matsedonskyi:2020kuy,Carena:2021onl,Biekotter:2021ysx,Bai:2021hfb,Matsedonskyi:2021hti,Chao:2021xqv,Chang:2022psj,Matsedonskyi:2022btb,Li:2023foh}, because it opens a window to realize baryogenesis at high energy scales. This can alleviate the tight phenomenological constraints~\cite{Baldes:2018nel,Glioti:2018roy,Carena:2021onl,Matsedonskyi:2022btb}. In this work, we study the high temperature $\mathbb{Z}_2$ symmetry breaking and the related GW from domain wall annihilation. We introduce a scalar  with $\mathbb{Z}_2$ symmetry and few of singlet fermions that interact with  scalar through a five-dimension operator. This five-dimension operator contributes a negative correction to effective potential, which leads to potential has a non-zero minimum at high temperature~\cite{Matsedonskyi:2020mlz}. As a result, the $\mathbb{Z}_2$ symmetry is broken at high temperature, and meanwhile the domain walls will form.\footnote{In Ref.~\cite{Ramazanov:2021eya}, the authors also conducted a similar study, but they breaks the $\mathbb{Z}_2$ symmetry at high temperature by introducing some new scalars with a negative coupling.}. In this scenario, the domain walls tension is temperature-dependent and the energy of domain walls drops faster than radiation~\cite{Ramazanov:2021eya,Babichev:2021uvl}. This is different from the standard domain walls where its tension is a constant and will overclose the Universe~\cite{Hiramatsu:2013qaa}. The domain wall annihilation can emit GW. Since the peak frequency of GW spectrum  is related to the energy scale of symmetry breaking, we can explore the high temperature symmetry breaking with GW experiments.

\par The paper is organized as follows. In Section~\ref{2}, we study the high temperature $\mathbb{Z}_2$ symmetry breaking and domain walls. In Section~\ref{3}, we discuss the GW from domain wall annihilation.  
In Section~\ref{5}, we summarize our results.

\section{High temperature $\mathbb{Z}_2$ symmetry breaking and domain walls}\label{2}
We first introduce a real scalar singlet $\phi$ with $\mathbb{Z}_2$ symmetry and few of singlet fermions $\psi$ that interact with  scalar through a five-dimension operator. The corresponding Lagrangian is~\cite{Matsedonskyi:2020mlz}
\begin{eqnarray}
\label{L}
\mathcal{L}\supset \frac{1}{2} (\partial \phi)^2-\frac{m_\phi}{2}\phi^2-\frac{\lambda}{4}\phi^4 -m_\psi \bar{\psi}\psi+\frac{\phi^2}{M_{\rm pl}}\bar{\psi}\psi,
\end{eqnarray}
where $m_\psi$ is the fermions bare mass and $M_{\rm pl}=(8\pi G)^{-1/2}=2.44 \times 10^{18}~\rm GeV$ is the reduced Planck mass. It is necessary to note that we just include one fermion in Eq.\eqref{L}. In this work, we introduce $N_f$ fermions with the same mass $m_\psi$. 
The scalar $\phi$ has a time-dependent and spatially homogeneous background field value. The equation of motion reads
\begin{eqnarray}
\label{EOM}
\ddot{\phi}+3H\dot{\phi}+\frac{d V_{\rm eff}}{d \phi}=0,
\end{eqnarray} 
where $V_{\rm eff}$ has the form\footnote{Here we  omit the term $\lambda T^2/2$ in thermal mass. This is reasonable because $\lambda\ll\kappa$ in our scenario, which is similar to the case shown in Ref.~\cite{Ramazanov:2021eya}.}
\begin{eqnarray}
\label{potential}
V_{\rm eff}(\phi, T)\approx -\frac{1}{2}\left( \frac{1}{3}\kappa T^2- m_\phi^2 \right)\phi^2+\frac{1}{4}\lambda\phi^4,
\end{eqnarray} 
here $\kappa$ is a dimensionless parameter
\begin{eqnarray}
\label{kappa}
\kappa=\frac{N_f m_\psi}{M_{\rm pl}}.
\end{eqnarray}
The thermal corrections in Eq.\eqref{potential} lead to the infinite temperature effective potential has a non-zero minimum at high temperatures. At the early stage, the  background  field $\phi$ is pinned to zero due to the large Hubble fraction. As the temperature drops, when the scalar thermal  mass becomes comparable to Hubble parameter, $\phi$ quickly roll down to the minimum of potential.  
Meanwhile, the $\mathbb{Z}_2$ symmetry is spontaneously broken and the scalar $\phi$ has equal probability to choose positive and negative values in different Hubble patches. The patches with different vacuum values are separated by domain walls. We estimate the temperature at domain walls formation as
\begin{eqnarray}
T_{i}=\sqrt{\frac{\kappa}{3g_{\ast}(T_i)}}M_{\rm pl}.
\end{eqnarray}
The tension of domain walls is
\begin{eqnarray}
\label{tension}
\sigma_{\rm wall}(T)=\frac{(2\kappa)^{3/2}}{9\sqrt{3}\lambda}T^3.
\end{eqnarray} 
It is necessary to note that  the above $T$-dependent tension $\sigma_{\rm wall}(T)\propto T^3$  is different from the standard domain walls tension where $\sigma_{\rm wall}=constant$. As a result, the energy density of  $T$-dependent domain walls in the scaling regime $\rho_{\rm wall} \propto T^5/T_i$  decays fast and  may avoid to overclose the Universe~\cite{Ramazanov:2021eya}. Specifically, in this work, in oder to avoid the domain wall problem, the following condition must be satisfied\footnote{We assume that the domain walls are formed at the radiation dominated stage.}~\cite{Ramazanov:2021eya}
\begin{eqnarray}
\label{tension1}
\frac{\rho_{\rm wall}}{\rho_{\rm rad}}\approx \frac{\kappa^2}{10 g_\ast(T_i)\lambda}\cdot \frac{T}{T_i}<1,
\end{eqnarray} 
here $\rho_{\rm rad}=\pi^2 g_\ast(T)T^4/30$ is the radiation energy density.

As the temperature of the Universe decreases, when the scalar thermal mass becomes comparable to its bare mass $\frac{1}{3}\kappa T^2\simeq m_\phi^2$, the $\mathbb{Z}_2$ symmetry will be restored. This process is a second order phase transition.

\section{GW from domain wall annihilation}\label{3}
In this section, we study the GWs emitted by domain wall annihilation. The spectrum of GW at the cosmic time $t$ is~\cite{Maggiore:1900zz,Maggiore:1999vm}
\begin{equation}
\Omega_{\rm gw}(t,f) = \frac{1}{\rho_{\rm tot}(t)}\frac{d\rho_{\rm gw}(t)}{d\ln f},
\end{equation}
where $\rho_{\rm tot}(t)=3H^2/M_{\rm pl}^2$ is the total energy of the Universe. $f=k/2\pi R(t)$ is the frequency corresponding to the comoving wavenumber $k$.
The numerical simulations gives the peak amplitude at the annihilation time of domain walls~\cite{Hiramatsu:2013qaa}
\begin{equation}
\Omega_{\rm gw}(t_{\rm ann})_{\rm peak}  = \frac{8\pi\tilde{\epsilon}_{\rm gw}G^2\mathcal{A}^2\sigma_{\rm wall}^2}{3H^2(t_{\rm ann})}. \label{Omega_gw_t_dec_peak}
\end{equation}
Here we assume that the production of GW  happens during the radiation dominated era. For the $T$-dependent domain walls, its annihilation time  is close to its formation time $t_{\rm ann}\simeq t_{i}$ as shown in Ref.~\cite{Ramazanov:2021eya}, and we will use this in the following text. The parameters $\tilde{\epsilon}_{\rm gw}= 0.7 \pm 0.4$ and $\mathcal{A}= 0.8 \pm 0.1$. Then we get the peak amplitude of GW at the present time $t_0$~\cite{Hiramatsu:2013qaa}
\begin{align}
\Omega_{\rm gw}h^2(t_0)_{\rm peak} &= \frac{\rho_{\rm gw}(t_0)h^2}{\rho_{\rm tot}(t_0)}  \nonumber\\
&= \Omega_{\rm rad}h^2\left(\frac{g_*(T_{i})}{g_{*0}}\right)\left(\frac{g_{*s0}}{g_{*s}(T_{i})}\right)^{4/3}\Omega_{\rm gw}(t_{i})_{\rm peak}, \label{Omega_gw_h2_t0}
\end{align}
where $\Omega_{\rm rad}h^2 = 4.15\times 10^{-5}$ is the density parameter of radiations today. $g_{*0} = 3.36$ and $g_{*s0} = 3.91$ are the effective relativistic degrees of freedom at the present time for the energy density and the entropy density, respectively. $h = H_0/100\,\mathrm{km}\cdot\mathrm{sec}^{-1}\mathrm{Mpc}^{-1}$ is the reduced Hubble parameter.
Finally we write the peak amplitude of GW as
\begin{eqnarray}
\Omega_{\rm gw}h^2(t_0)_{\rm peak}\approx 9 \times 10^{-14}\cdot\frac{\kappa^4}{\lambda^2}\cdot\left(\frac{100}{g_\ast(T_i)}\right)^{7/3}.
\end{eqnarray}
The  peak frequency  of GW is determined by $f_{\rm gw}(t_i)_{\rm peak}\simeq H(T_i)$. Considering the redshift factor $a(t_i)/a(t_0)$, we can get the peak frequency at the present time  
\begin{eqnarray}
f_{\rm gw}(t_0)_{\rm peak}\approx 2 \times 10^{11}~ {\rm Hz}\cdot\sqrt{\kappa}\cdot\left(\frac{100}{g_\ast(T_i)}\right)^{1/3}.
\end{eqnarray}
For more details about the GW form domain wall annihilation, one can refer to the review paper~\cite{Hiramatsu:2013qaa}.

\begin{figure}[tbp]
	\begin{center} 
	\includegraphics[width=12.8cm]{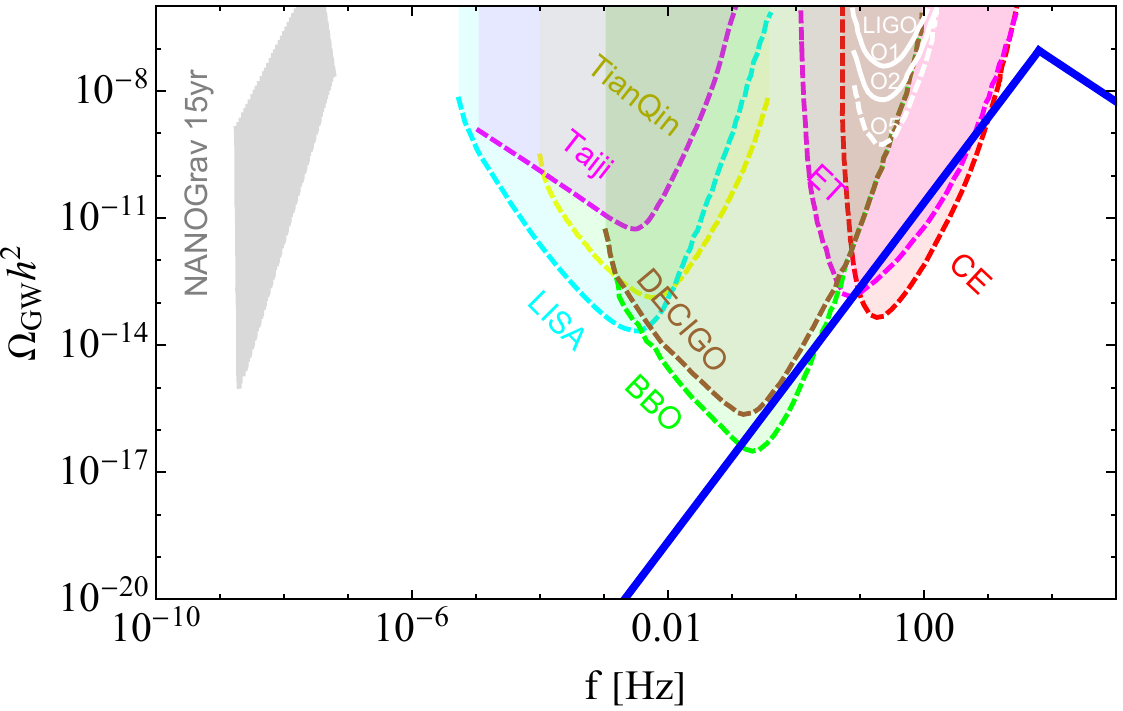}
	\end{center} 
	\caption{Sensitivities of GW detectors and the GW spectrum (blue solid line) from  domain wall annihilation for $\kappa\sim 10^{-15}$ and $\lambda\sim 10^{-33}$. The domain walls are formed at $\mathcal{O}(10^{9})~ \rm GeV$. 
  }
	\label{GW1}
\end{figure}

\begin{figure}[tbp]
	\begin{center} 
	\includegraphics[width=12.85cm]{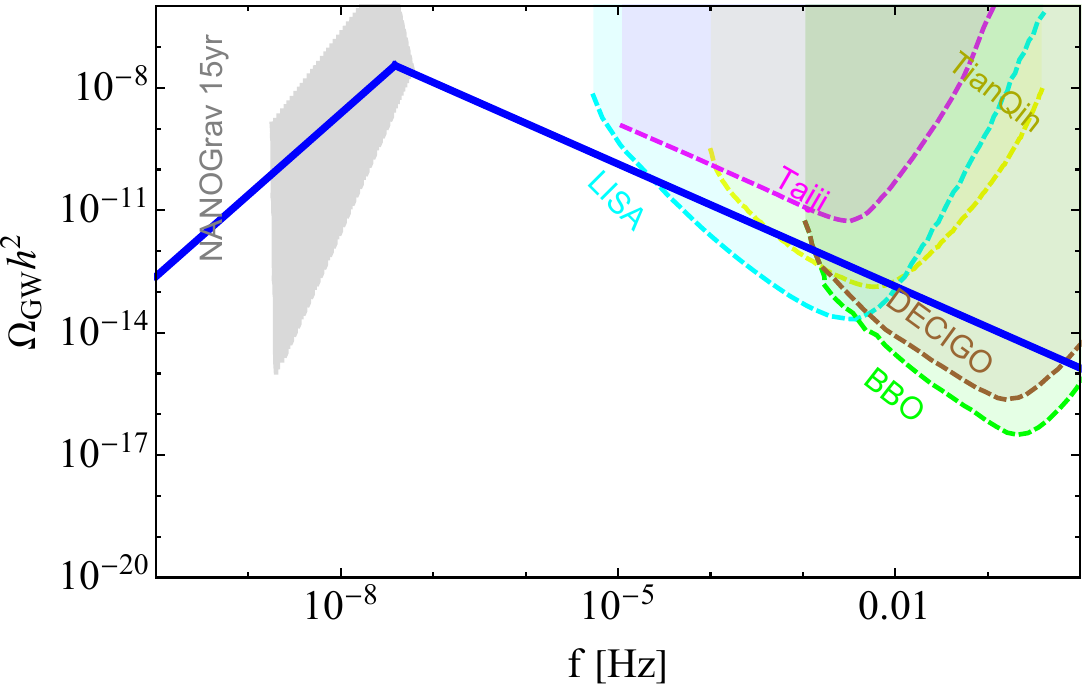}
	\end{center} 
	\caption{Sensitivities of GW detectors and the GW spectrum (blue solid line) from  domain wall annihilation for $\kappa \sim 10^{-38}$ and $\lambda \sim 10^{-78}$. The domain walls are formed at $\mathcal{O}(10)~ \rm MeV$ and the peak frequency of GW spectrum around $\mathcal{O}(10^{-8})~ \rm Hz$. }
	\label{GW2}
\end{figure}

Figure.~\ref{GW1} shows the sensitivities of GW detectors and the GW spectrum from domain wall annihilation for $\kappa\sim10^{-15}$ and $\lambda\sim10^{-33}$. The blue solid line represents the GW energy spectrum. The  dashed lines are the power-law integrated sensitivity curves for the Taiji \cite{Guo:2018npi}, LISA \cite{Audley:2017drz,Breitbach:2018ddu}, TianQin \cite{Mei:2020lrl}, BBO \cite{Cutler:2005qq}, DECIGO \cite{Kudoh:2005as,Musha:2017usi}, ET \cite{Hild:2010id,Punturo:2010zz}, CE \cite{LIGOScientific:2016wof} and LIGO \cite{LIGOScientific:2014qfs,LIGOScientific:2019vic}. The gray shaded region stands for the NANOGrav 15-year signal~\cite{NANOGrav:2023gor}. We find the domain walls are formed at $\mathcal{O}(10^{9})~ \rm GeV$ and the GW produced by domain wall annihilation is expected to be observed by BBO, CE and ET.

Next, let us discuss the relationships between this scenario and NANOGrav signal~\cite{NANOGrav:2023gor}. Figure.~\ref{GW2} shows the GW spectrum from the domain wall annihilation for $\kappa\sim10^{-38}$ and $\lambda \sim 10^{-78}$. We find the domain walls are formed at $\mathcal{O}(10)~ \rm MeV$ and the peak frequency of GW spectrum $f\sim10^{-8}~\rm Hz$, which is expected to explain NANOGrav the 15-year signal. In addition, the GW at high frequency region is also expected to be observed by LISA, TianQin, DECIGO and BBO.

\section{Summary and conclusions}\label{5}
In this work, we have studied the high temperature $\mathbb{Z}_2$ symmetry breaking and the GW from domain wall  annihilation. We breaks the $\mathbb{Z}_2$ symmetry at high temperature  by introducing few of singlet fermions which interact with scalar through a five-dimension operator. At the early stage, the scalar is pinned at symmetric phase due to the large Hubble fraction until its thermal  mass becomes comparable to the Hubble parameter. After that, the scalar begins to quickly roll down to the minimum of potential. This is similar to the thermal misalignment \cite{Batell:2021ofv}. At the same time,  $\mathbb{Z}_2$ symmetry is spontaneously broken and the domain walls will form. In this scenario, the tension of the domain wall is temperature-dependent and the energy density of domain wall drops faster than radiation, which avoids the domain wall problem.  With the decrease of temperature, the thermal mass of scalar becomes gradually small and finally the $\mathbb{Z}_2$ symmetry will be restored.

The domain wall annihilation can emit GW. Since the peak frequency of GW spectrum is related to the energy scale of $\mathbb{Z}_2$ symmetry breaking, we can explore the high temperature symmetry breaking with GW experiments. The characteristics of GW spectrum also depend on parameters. In our work, the self-interaction coupling  of scalar i.e. $\lambda$  ~is tiny that is similar to the case shown in Ref.~\cite{Ramazanov:2021eya}.  We find that: 

(a).~If parameters $\kappa\sim 10^{-15}$ and $\lambda \sim 10^{-33}$, the $\mathbb{Z}_2$ symmetry is broken at $\mathcal{O}(10^9)~\rm GeV$ and the peak frequency of GW emitted by domain wall annihilation is around $\mathcal{O}(10^3\sim10^4)~\rm Hz$. It is expected to be observed by BBO, CE and ET. 

(b).~If parameters $\kappa\sim 10^{-38}$ and $\lambda \sim 10^{-78}$, the $\mathbb{Z}_2$ symmetry is broken at $\mathcal{O}(10)~\rm MeV$ and the peak frequency of GW is around $\mathcal{O}(10^{-8})~\rm Hz$.  
This may provide an explanation for the NANOGrav signal~\cite{NANOGrav:2023gor}.

Finally, we note that since $\kappa \sim  10^{-38}$  in case~(b) is very tiny, some ultralight fermions with mass around $\mathcal{O}(10^{-11})~\rm eV$ may appear (Please see Eq.~\eqref{kappa}).  Those ultralight fermions may serve as \emph{ultralight fermionic dark matter}  which is heavier than  $\sim 10^{-14}$ eV as shown in Ref.~\cite{Davoudiasl:2020uig}. We leave this for future research.


\section*{Acknowledgments}
The author thanks Ying-Quan Peng for valuable discussions. This work is supported by Scientific Research Foundation of Inner Mongolia University under grant No.10000-22311201/036.

\bibliographystyle{apsrev4-1}
\bibliography{Z2GW}

\end{document}